\documentclass[12pt]{article}\usepackage[]{graphicx}\usepackage[]{color}
\makeatletter
\def\maxwidth{ %
  \ifdim\Gin@nat@width>\linewidth
    \linewidth
  \else
    \Gin@nat@width
  \fi
}
\makeatother

\definecolor{fgcolor}{rgb}{0.345, 0.345, 0.345}

\usepackage{framed}
\makeatletter
 {\par\unskip\endMakeFramed%
 \at@end@of@kframe}
\makeatother

\definecolor{shadecolor}{rgb}{.97, .97, .97}
\definecolor{messagecolor}{rgb}{0, 0, 0}
\definecolor{warningcolor}{rgb}{1, 0, 1}
\definecolor{errorcolor}{rgb}{1, 0, 0}

\usepackage{alltt}
\usepackage{amsmath, amssymb, booktabs, setspace, bm}
\usepackage{graphicx,psfrag,epsf}
\usepackage{enumerate}
\usepackage{natbib}
\usepackage{url}

\newcommand{\blind}{0}

\IfFileExists{upquote.sty}{\usepackage{upquote}}{}
\begin{document}

\def\spacingset#1{\renewcommand{\baselinestretch}%
{#1}\small\normalsize} \spacingset{1}


\if0\blind
{
  \title{\bf Enriching Students' Conceptual Understanding of Confidence Intervals: An Interactive Trivia-based Classroom Activity}
  \author{Xiaofei Wang\hspace{.2cm}\\
    Department of Statistics, Yale University\\ \\
    Nicholas G. Reich\hspace{.2cm}\\
    Department of Biostatistics and Epidemiology, \\
    University of Massachusetts at Amherst\\ \\
    Nicholas J. Horton\hspace{.2cm}\\
    Department of Mathematics and Statistics, Amherst College}
  \maketitle
} \fi

\if1\blind
{
  \bigskip
  \bigskip
  \bigskip
  \begin{center}
    {\LARGE\bf Enriching Students' Conceptual Understanding of Confidence Intervals: An Interactive Trivia-based Classroom Activity}
\end{center}
  \medskip
} \fi

\bigskip
\begin{abstract}
Confidence intervals provide a way to determine plausible values for a population parameter. They are omnipresent in research articles involving statistical analyses. Appropriately, a key statistical literacy learning objective is the ability to interpret and understand confidence intervals in a wide range of settings. As instructors, we devote a considerable amount of time and effort to ensure that students master this topic in introductory courses and beyond.  Yet, studies continue to find that confidence intervals are commonly misinterpreted and that even experts have trouble calibrating their individual confidence levels. In this article, we present a ten-minute trivia game-based activity that addresses these misconceptions by exposing students to confidence intervals from a personal perspective. We describe how the activity can be integrated into a statistics course as a one-time activity or with repetition at intervals throughout a course, discuss results of using the activity in class, and present possible extensions.

\end{abstract}

\noindent%
\vfill

\newpage
\spacingset{1.45} 
\section{Introduction}
\label{sec:introduction}

Confidence intervals are one of the most commonly used statistical methods to summarize uncertainty in parameter estimates from data analyses. However, both the formal concept of and intuition behind confidence intervals remain elusive to many students and data analysts. In the case of a 95\% confidence interval for a proportion for a given sample size, one textbook definition is ``95\% of samples of this size will produce confidence intervals that capture the true proportion'' \citep{book:1321656}. Another textbook definition reads ``A plausible range of values for the population parameter is called a confidence interval'' and later specifies for 95\% confidence ``Suppose we took many samples and built a confidence interval [for the population mean] from each sample...then about 95\% of those intervals would contain the actual mean'' \citep{diez2012openintro}.
Studies have found that many students struggle with understanding definitions like these even after completing coursework at various levels \citep{Fidler:2005ur,Kalinowski:2010tq,Kaplan:2010vd}.

In addition to a lack of intuition about the definition of a confidence interval, researchers have documented overconfidence in quantitative assessments of uncertainty in students and experts alike \citep{Alpert:A8vJTwl8,soll2004,jorgensen2004,cesarini2006}.
In this context, overconfidence has been defined as individuals having ``excessive precision'' in their beliefs about particular facts \citep{moore2008}.
One study showed that when subjects were asked to provide 98\% confidence intervals relating to numeric facts, only 57.4\% of their intervals captured the true value \citep{Alpert:A8vJTwl8}. This display of overconfidence is not limited to students and novices, but is also exhibited by field experts \citep{hynes1976reliability,christensen1981physicians}.

The textbook definition of a confidence interval can be challenging to understand, and various pedagogical tools have been proposed to assist with comprehension.
Some teachers have documented success by teaching confidence intervals with bootstrapping techniques \citep{Maurer:2015vq} or with visualizations of simulated samples \citep{Cumming:2007cl,Hagtvedt:2008jv}. \cite{behar2013twenty} presented one analogy useful in understanding 95\% confidence intervals: ``It is like a person who tells the truth 95\% of the time, but we do not know whether a particular statement is true or not.''
Providing students with an opportunity to calibrate their intuitive understanding of what it means to be, say, 90\% confident about a fact through short repetitive activities may enhance their ability to understand the strength of conclusions from a data analysis.
In this manuscript, we describe an activity that aims to solidify an intuitive understanding of what it means to be ``90\% confident''.



Motivated by \cite{Alpert:A8vJTwl8}, our short interactive classroom activity ties together formal concepts of confidence intervals to tangible information and questions about facts that students may or may not have some familiarity with and interest in. In brief, the instructor reads off ten trivia questions with specific quantitative answers. Students are asked to provide their answer to each question in the form of a 90\% confidence interval (rather than just providing a point estimate). After all the trivia questions are read, the correct answers are revealed and students calculate the number of intervals that capture the correct answers within their intervals.

By providing immediate feedback and fostering a gently competitive spirit in the classroom, this activity incentivizes students to engage with the central conceptual challenges of confidence intervals as described in the GAISE College Report \citep{GAISEcollege}.

The questions should be appropriate for the audience, and can be about topics that are timely and relate to cultural memes (for example, celebrities, sports teams, and TV shows). However, it is important to emphasize that this is not a test of knowledge, but rather of how well one knows and is able to quantify the limits of their knowledge.  Specifically, if students are not familiar with a particular topic, they can (and should) respond with a wide interval.
Assessing personal confidence interval coverage rates provides immediate feedback about how well-calibrated a student's confidence level is.

The activity is simple, requires little preparation, and is appropriate for students at all different levels and backgrounds, including undergraduate and graduate students. The activity requires only a list of 10 questions with quantitative answers and can be used as early as in an introductory statistics course when the students are first exposed to confidence intervals. In an upper level course, the activity can be used early on for review. Given the simplicity of the exercise, this activity can be utilized in a classroom of twenty students or two hundred students with little modification.


In Section \ref{sec:activity}, we describe the activity procedure and necessary preparation. Section \ref{sec:results} presents anecdotal and numerical evidence relating to the instructional value of the activity. Finally, in Section \ref{sec:discussion}, we present a discussion relating to the effectiveness of the activity.

\section{Activity} 
\label{sec:activity}

\subsection{Overview}
Before class, the instructor needs to prepare a series of ten or more trivia questions that each have specific quantitative answers. There are many sources for ideas. Trivia questions about the college or university may be popular (e.g. how many official student organizations are listed on the school website?). We have used the board game Wits and Wagers\footnote{Crapuchettes, D. and N. Heasley (2005). \underline{Wits and Wagers}. Bethesda, MD: North Star Games.} as one source of questions. The questions should ideally cover a broad range of topics so that most students will not know the answers exactly, but that educated guesses are possible. For example, a question such as ``In what year was the Declaration of Independence signed?'' does have a quantitative answer but would be a poor choice because most students would be able to answer 1776 without hesitation. Table \ref{tbl:example} gives a series of example questions that might be used for this activity. These questions were obtained by one of the authors via the Wits \& Wagers: Trivia Party iOS app\footnote{\url{https://itunes.apple.com/us/app/wits-wagers-trivia-party/id637929057?mt=8}}. Appendix A contains a list of thirty additional questions that can be used for this activity.

We now outline the steps of the activity assuming that ten questions have already been selected for use.
\begin{enumerate}
  \item Provide students with a blank sheet of paper and ask the students to make a numbered list from 1 to 10, leaving room for ten answers.
  \item Tell students that they will hear a series of ten questions, each one having a numeric answer. Instead of writing down a specific value to answer each question, students should provide their answer in the form of a 90\% confidence interval, i.e. (lower bound, upper bound). Optionally, provide an example question and a corresponding interval (not scored) so that students understand the format of the exercise.
  \item Read each of the ten trivia questions, pausing for sufficient time (typically 20-30 seconds) in between questions to give students a chance to jot down their answer.
  \item After all questions are read, explain the scoring process; as you read the answers to each question, the students should give themselves a point for every interval that captures the correct numeric response. (Alternatively, you might arrange for students to switch answers with a neighbor to have them do scoring for one other.)
  \item Review the answers with the students and ask students to tally up their score, out of ten. (This process is often high-energy with lots of reaction from the students.)
  \item Obtain and display a distribution of student scores. If there are a small number of students, a show of hands suffices for data collection. Otherwise an online survey or a clicker poll might be useful for collecting the information. The instructor might summarize results in a sketched stem-and-leaf plot or histogram.
  \item Discuss the results as a group. A proposed starting point could be ``what might you expect the results to look like?'' or ``what do the observed results indicate about the over- or under-confidence of the class as a whole?''
\end{enumerate}

In our experience, students tend to score low the first time that they do this exercise (see Section~\ref{sec:results}); they tend to be overconfident in their answers. The last step of the exercise gives students an opportunity to reflect on both what it means to be overconfident and what the source of that miscalibration might be. As part of the discussion, one might ask what the expected score would be for a student who indeed provided 90\% confidence intervals (nine out of ten). More advanced classes could compare the distribution to repeated draws from a Binomial distribution with $N=10$ and various probabilities of success. Some students may have scored poorly due to an overestimation of how well they know the subject areas in question. Other students may have been overeager to play a competitive trivia game and may have overlooked the specification of the 90\% confidence level that should be considered as part of their answers. For all students, this activity serves to bridge the gap between a textbook definition of confidence intervals to a more intuitive understanding of confidence and uncertainty.

\begin{table}
  \centering
  \begin{tabular}{@{}rp{5.5in}r@{}}
  \toprule
  & \textbf{Question} & \textbf{Answer}\\
  \midrule
  1 & How old was actress Drew Barrymore when her first feature film was released? & 5\\
  2 & How many cards are in an original Uno deck? & 108\\
  3 & The first item ever sold on eBay was a broken laser pointer. In dollars, how much did it sell for? & 14\\
  4 & According to the 1994 film Forrest Gump, what was Gump's IQ? & 75\\
  5 & In years, how long did it take Michelangelo to paint the ceiling of the Sistine Chapel? & 4\\
  6 & In months, how long did it take Apple to sell 100 million iPods? & 66\\
  7 & How many episodes aired of the TV sitcom Friends, not including specials? & 236\\
  8 & In what year was the first modern Summer Olympics held? & 1896\\
  9 & In miles, how long is the main wall of the Great Wall of China, excluding all of its secondary branches? & 4163\\
  10 & How many states were part of the United States in 1860? & 33\\

  \bottomrule
\end{tabular}
\caption{Example Trivia Questions}\label{tbl:example}
\end{table}

For the convenience of the instructor, we have included a one-sheet summary of the activity with example questions, scoring, and discussion questions in the Supplementary Materials. This could be printed out to use in class.

\subsection{Variations and Evaluation}
We note that when exactly ten questions are utilized, students can `cheat' the activity by giving nine extremely wide intervals and one obviously wrong interval. In doing so, they can almost definitely guarantee a score of nine out of ten. One way to fix this loophole is to modify the number of questions used. For instance, students might be asked 15 questions but only a random selection of 10 are used for scoring. Additionally, the activity could be extended by asking students to think about and propose a metric that could identify this kind of cheating; for example, a mean and standard deviation could be calculated from all intervals for a question, and students having multiple interval widths more than some $z$-scores above the mean could be flagged as potential cheaters.

The proposed activity can take many forms and be used for different purposes. As a group, the authors have used this activity with both graduate and undergraduate students, at one or more times during the semester, and with slightly different learning goals in mind. Two of the authors first use the activity in an undergraduate introductory statistics course during the class period immediately following the introduction of confidence intervals with a follow-up towards the end of the semester. Another has used it repeatedly throughout the semester (using different questions each time) in a graduate-level course on regression that is a required course for statistics Masters students. In this course, the goal of repeating the activity is to illustrate the challenges of calibrating uncertainty, i.e. avoiding ``excessive precision'', and understanding what it means to be 90\% confident about a fact. In each of the courses described above, the instructors also employ small simulation-based exercises to illustrate the properties of confidence intervals in a more formal statistical sense.

Apart from using this activity for understanding confidence intervals, the act of producing intervals and then validating whether the answers fall within those intervals serves as a way of calibrating one's level of confidence. We believe students stand to benefit from participating in this exercise several times over the course of a semester, of course, on different trivia questions. In our experience, students score better with each additional iteration (see Section \ref{sec:results}). Such incremental improvements suggest that students learn to better quantify their levels of uncertainty with practice.

In the graduate-level course where the activity was repeatedly used, the  Comprehensive Assessment of Outcomes in a First Statistics course, or the CAOS test, was used to evaluate overall conceptual understanding of statistics \citep{delmas2007}. Specifically, the CAOS pre- and post-tests were administered at the beginning and end of the semester. We include a comparison of pre- and post-test scores from confidence interval-related questions in the Supplementary Materials.

\section{Results} 
\label{sec:results}

Over the course of one semester, the three authors used this activity two, two, and five times in their respective classes. We now describe the results with data consisting of student scores from these three classes. The data collection, management, and analysis were approved by the University of Massachusetts at Amherst Institutional Review Board (\#2014-2051) and Amherst College Institutional Review Board (\#16-008).

The authors that used the activity twice in a semester (Horton and Wang) did so in undergraduate introductory statistics courses with 24 and 27 students, respectively. The first time that the activity was used (Iteration 1) was at the beginning of the lesson on confidence intervals, during Week 7 of a 13-week semester. Students were asked to pre-read the chapter on confidence intervals for a single proportion prior to coming to class; class began with the activity without discussion of the reading material. The second time that the activity was used (Iteration 2) was during Week 11 after students had seen confidence intervals used in two-sample settings and in linear regression. Table~\ref{tbl:data_summary} shows that the students performed poorly on their first attempt, averaging 4.5 and 3.3 in the two classes, and improved significantly on the second attempt, averaging 5.7 and 5.8. Identifying information was not tracked during either of the iterations, so it is not possible to track individual-level improvements.

The third author (Reich) used this activity in a graduate-level multivariate regression course five times spread approximately every two weeks throughout the semester and recorded results from each round of the activity. Students entered their data into a shared spreadsheet, with a unique student identifier across all rounds. Out of 14 students total, 11 participated in all five rounds of the activity, while two participated in all but the first round and one other participated in the third to fifth rounds. (One student declined to have their scores used in the subsequent data analysis.)

\begin{figure}
\centering
  \includegraphics[width=0.7\textwidth]{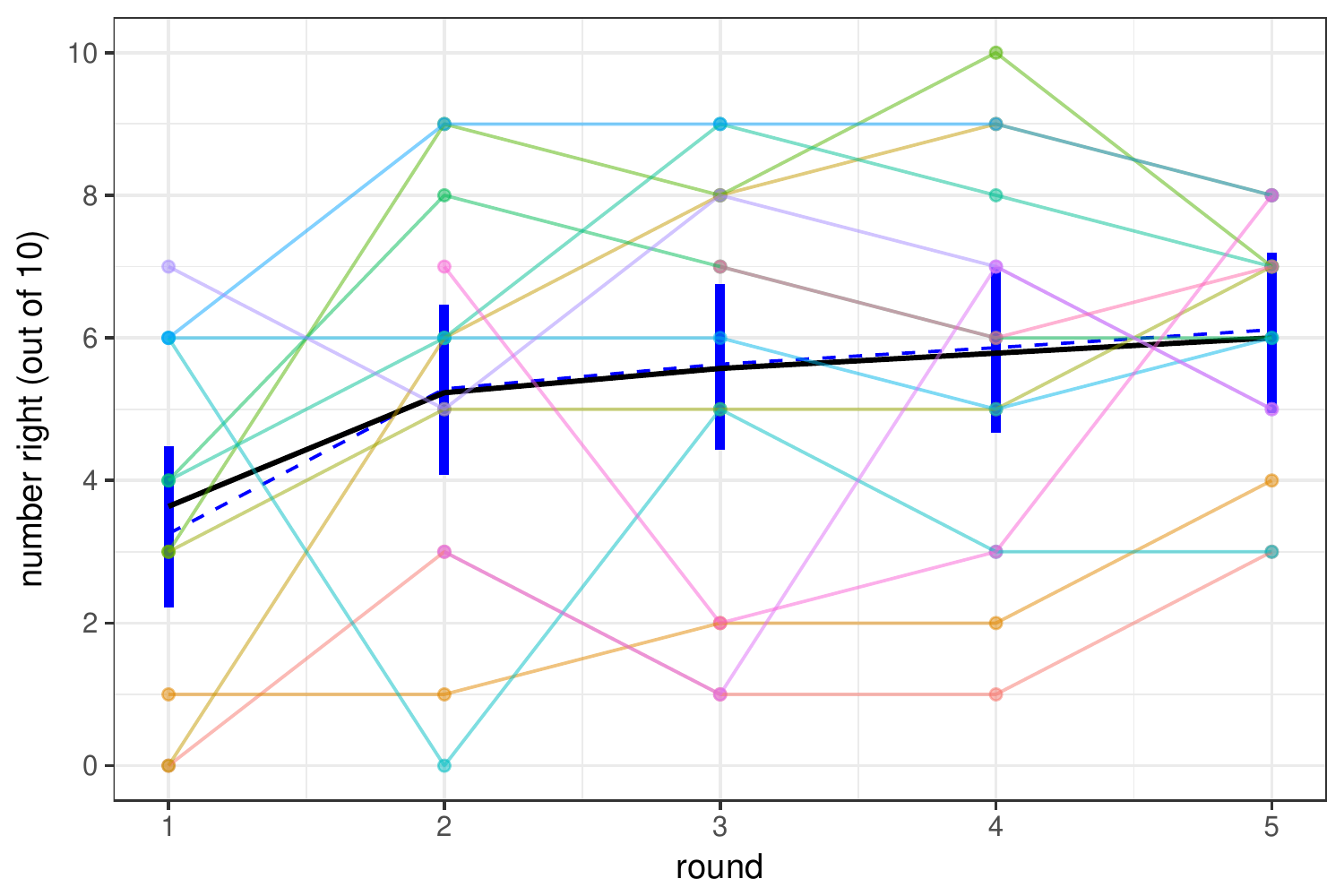}
  \caption{Student scores from five iterations of the activity ($n=14$, shown in lighter colors, color-coded by student). The solid black line indicates the mean score across all students. The dashed blue line indicates the model-estimated posterior median number of questions answered correctly, with corresponding 90\% confidence intervals shown in vertical blue lines.}
  \label{fig:analysis}
\end{figure}

The longitudinal data from this class are plotted in Figure~\ref{fig:analysis}, along with the overall mean score per iteration and model estimates of per-iteration success rates. The average number of intervals that covered the true values in the first and fifth iterations were 4 and 6.5, respectively. While there is substantial within-student variation (as shown by the student-specific lines), the overall trend of the median number of intervals that cover the truth increased with repetition of the activity. Over the five iterations, only in one instance was there an individual with all ten intervals covering the truth.

To enable inference on per-iteration improvement while accounting for within-subject correlations, we fit a logistic mixed-effects model with a random intercept for each individual student and fixed effect dummy variables for iteration. We used the {\tt rstanarm} package \citep{rstanarm} in R version 3.3.2 (2016-10-31) \citep{rcoreteam} to fit the model with a Bayesian Markov Chain Monte Carlo algorithm, using standard weakly informative priors for the coefficients and variance parameters. While accuracy increased with each iteration, the biggest incremental score gain occurred between the first and second iterations, with an improvement of 1.5 to 2 questions answered correctly (Table~\ref{tbl:data_summary}). Accuracy increased more slowly between iterations 2 through 5, although an average improvement of almost a single correct question was observed across these four iterations. Comparing the 90\% posterior credible intervals (CI) for the number of questions answered correctly per iteration, only rounds 4 (90\% CI: (4.7, 7)) and 5 (90\% CI: (4.9, 7.2)) showed significant increases in accuracy over iteration 1 (90\% CI: (2.4, 4.4)); the credible intervals for iterations 2 and 3 still overlapped with the credible interval for iteration 1.

\begin{table}[ht]
\centering
\begin{tabular}{@{}lccccccccc@{}}
  \toprule
  & \multicolumn{2}{c}{Horton} & \multicolumn{2}{c}{Wang} & \multicolumn{5}{c}{Reich}\\
\cmidrule(lr){2-3} \cmidrule(lr){4-5}  \cmidrule(lr){6-10}
  Iteration & 1 & 2 & 1 & 2 & 1 & 2 & 3 & 4 & 5 \\
  \midrule
n & 21 & 21 & 26 & 19 & 11 & 13 & 14 & 14 & 14 \\
  mode & 5 & 3 & 1 & 4 & 6 & 6 & 8 & 3 & 7 \\
  median & 5.0 & 6.0 & 3.0 & 6.0 & 4.0 & 6.0 & 6.5 & 6.0 & 6.5 \\
  mean & 4.5 & 5.7 & 3.3 & 5.8 & 3.6 & 5.2 & 5.6 & 5.8 & 6.0 \\
  sd & 1.8 & 2.1 & 2.3 & 2.2 & 2.5 & 2.8 & 3.0 & 2.8 & 1.8 \\
  \midrule
  estimated mean    &  &  &  &  &
  3.3 &
  5.3 &
  5.6 &
  5.9 &
  6.1\\
  estimated 90\% CI &  &  &  &  &
  2.2-4.5 &
  4.1-6.5 &
  4.4-6.8 &
  4.7-7 &
  4.9-7.2 \\

   \bottomrule
\end{tabular}
\caption{Summary Statistics of Activity By Class and Iteration. The last two rows represent model-based estimates.}\label{tbl:data_summary}
\end{table}

\section{Discussion} 
\label{sec:discussion}

Confidence intervals play a key role in inferential thinking. We have described an interactive activity to help solidify students' understanding of confidence intervals. The activity is attractive because it requires no technology, is interactive and applicable to a wide variety of students, and fun, in the spirit of friendly competition between students. The activity can help reinforce statistical thinking and help students recognize overconfidence. However, it should be noted as a limitation of the activity that the intervals requested of the students are somewhat removed from the typical notion of a confidence interval; students are not computing standard errors and point estimates but are rather relying on their own judgement. Nonetheless, because the activity does not involve any calculations, which tends to be the typical entry point for teaching confidence intervals, it can be particularly helpful for promoting an intuitive understanding. Students with very little mathematical background, such as graduate students from non-quantitative fields, may find this activity a helpful way to understand confidence intervals without the distraction of the mathematical formulae.

It is interesting to note that while scores improve over repeated iterations of the activity, perfect calibration of confidence was not attained within the course of a single semester. Even with five iterations of the activity, student scores plateau between 5 to 6 correct answers out of 10, well below the desired level of 9. Moreover, students in the classes that only used the activity twice in a semester seemed to do about as well in the second iteration as those who had a third, fourth, or fifth try. This observation suggests that overconfidence is challenging to overcome. Examining whether the same achievement gap holds if students were asked to produce confidence intervals at different levels (i.e. 50\% or 70\%) could provide information on whether this is a problem inherent to individual's judgment uncertainty, or whether 90\% is a particularly tricky level to achieve.

We relied on the CAOS test (details provided in the Supplementary Materials) to provide a quantitative evaluation of student's understanding of confidence intervals. The data suggested improvement in students' overall understanding of the interpretation of confidence intervals, although the sample size was small (15 students took both tests) and improvements in understanding could also be attributed to other content in the course.

With the continued increase of written content by ``data journalists'' in the mass media (often including some technical measures of uncertainty), the value of accurate intuition about uncertainty will grow for not just students and practitioners of statistics but for the general public as well. The results shown in this manuscript serve as proof-of-concept that asking students to create confidence intervals for quantities that are of potential interest can foster a more intuitive understanding of confidence and uncertainty. Our experience is that students are consistently surprised by the results of this exercise, leading to ``teachable moments'' when students grasp how and why their interval coverage was too low. Therefore, activities such as the one presented here and others may serve an important role in creating informed consumers of modern data-driven content, whether journalistic or academic, by arming them with tools to interpret quantitative results.

\section{Acknowledgements} 
\label{sec:acknowledgements}
First we would like to thank the Associate Editor and two anonymous reviewers who provided very helpful feedback on the first iteration of this manuscript. We also thank the students who participated in these activities. Finally, we thank Eric W. Bright, CFA, who introduced NGR to a version of this activity.

\newpage
\bibliographystyle{dcu}
\bibliography{refs}
\newpage

\section*{Appendix A: 30 Questions} 
\label{sec:appendix_a_30_questions}
In this section, we provide thirty more questions in addition to the ones presented in Table~\ref{tbl:example} for convenience. The first ten questions are adapted from \cite{confidentdecision} (reprinted in \cite{supercrunchers}, among other places), the next nine are taken from the Wits and Wagers board game, and the rest are written up by the authors.

\begin{table}[!ht]
  \centering
  \begin{tabular}{@{}rp{5.5in}r@{}}
  \toprule
  & \textbf{Question} & \textbf{Answer}\\
  \midrule
  1 & At what age did Martin Luther King die? & 39 \\
  2 & In miles, how long is the Nile River? & 4,187\\
  3 & In 2016, how many countries are part of OPEC? & 12 \\
  4 & How many books were part of the Old Testament? & 39\\
  5 & What is the diameter of the moon in miles? & 2,160\\
  6 & In pounds, what is the weight of an empty Boeing 747? & 390,000\\
  7 & In what year was Wolfgang Amadeus Mozart born? & 1756\\
  8 & What is the gestation period, in days, of an Asian elephant? & 645\\
  9 & In miles, what is the air distance from London to Tokyo? & 5,959\\
  10 & How deep, in feet, is the deepest known point in the oceans? & 36,198\\
  11 & In what year did an actress first earn \$1 million for a movie role? & 1963\\
  12 & In feet, how tall is The Statue of Liberty including the pedestal? & 305\\
  13 & On average, how many quarts of ice cream did an American eat in the year 2012? & 20\\
  14 & In pounds, what was the weight of the heaviest domesticated cat ever recorded? & 46.81\\
  15 & How many points did Michael Jordan average per game during his NBA career? & 30.12\\
  16 & In years, how long does the US \$1 bill stay in circulation? & 4.8\\
  17 & How many times could Rhode Island fit into the land area of Alaska? & 423\\
  18 & How many inventions did Thomas Edison patent during his lifetime? & 1,093\\
  19 & In what year was a woman first appointed to the Supreme Court? & 1981\\
  20 & In 2012, how many customers frequented Disneyland per day on average? & 44,000\\
  21 & As of the end of year 2015, the New York City subway system was constituted of how many subway stations? & 469\\
  22 & How many James Bond films are there total as of 2016? & 26\\
  23 & What is the total height of Yosemite Falls, the highest water fall in Yosemite National Park? & 2,425\\
  24 & How many moons does Saturn have? & 62\\
  25 & As of the end of the 2015-2016 season, how many different basketball teams have won at least one NBA Championship? & 17\\
  26 & In degrees Fahrenheit, what is the average temperature at the North Pole in summer? & 32\\
  27 & If Earth was hollow, how many moons could fit inside it? & 50\\
  28 & How many times did Ron's name appear in all of the Harry Potter books? &5,784\\
  29 & In millions, how many copies of Adele's album '25' sold in the U.S. during its debut week? & 3.38\\
  30 & How many goals were scored during the 2016 soccer tournament Copa America? & 91\\
  \bottomrule
\end{tabular}

\end{table}
\subsection*{Sources for questions 20 to 30}
\footnotesize
\begin{enumerate}
\setcounter{enumi}{19}
\item \url{http://www.latimes.com/business/la-fi-disneyland-crowds-20150519-story.html}
\item \url{http://web.mta.info/nyct/facts/ridership/}
\item \url{https://en.wikipedia.org/wiki/List_of_James_Bond_films}
\item \url{https://en.wikipedia.org/wiki/Yosemite_Falls}
\item \url{https://en.wikipedia.org/wiki/Moons_of_Saturn}
\item \url{http://www.landofbasketball.com/championships/summary_of_winners.htm}
\item\url{http://www.landofbasketball.com/championships/summary_of_winners.htm}
\item \url{http://www.astronomy.org/programs/moon/moon.html}
\item \url{http://www.moviefone.com/2011/07/13/harry-potter-numbers-trivia/}
\item \url{http://www.billboard.com/articles/columns/chart-beat/6777959/adele-25-historic-chart-numbers}
\item\url{https://en.wikipedia.org/wiki/Copa_Am\%C3\%A9rica_Centenario}
\end{enumerate}

\newpage

\begin{center}
Supplementary Materials
\end{center}

\section*{Activity Printout}
\singlespacing
The first two pages serve as a convenient one-sheet (front and back) summary of the activity that can be printed out for use in class. 
\subsection*{Activity Outline} 
Activity lasts approximately 10-15 minutes. Steps follow:
\begin{enumerate}
  \item Each student starts out with a sheet of paper with numbered list from 1 to 10, leaving room for 10 answers. 
  \item You will read aloud 10 trivia questions each with a numeric answer. Instruct students to write down a 90\% confidence interval in response to each question. 
  \item Read each of the trivia questions. 
  \item Review answers to the trivia questions. Students should score themselves: 1 point for every interval that captures the correct numeric answer. 
  \item Ask students to tally up their score out of 10.
  \item Obtain and visualize a distribution of student scores.
  \item Discuss the results as a group.
\end{enumerate}
\subsection*{Possible Discussion Questions} 
\begin{enumerate}
  \item If a student provided 90\% confidence intervals in all ten cases, how many points would we expect him/her to score?
  \item If \textit{every} student provided 90\% confidence intervals in all ten cases, what would a histogram of scores look like for the class?
  \item Examining the histogram (or stem-and-leaf plot) of scores from our class, do you think we were overconfident or underconfident? 
  \item How might we, as a class, do better at this exercise?
\end{enumerate}

\subsection*{Sample Questions, Answers, and Scoring} 
\label{sub:sample_questions_and_answers}
In the table below, we reiterate the questions/answers presented in Table~1 and present sample responses for each. Responses that are \textbf{bolded} earn one point.
\begin{center}
  \begin{tabular}{@{}rp{4in}rr@{}}
  \toprule
  & \textbf{Question} & \textbf{Answer} & \textbf{Sample Response}\\
  \midrule
  1 & How old was actress Drew Barrymore when her first feature film was released? & 5 & $\bm{(1, 7)}$\\
  2 & How many cards are in an original Uno deck? & 108 & $(52, 104)$\\
  3 & The first item ever sold on eBay was a broken laser pointer. In dollars, how much did it sell for? & 14 & $\bm{(3,200)}$\\
  4 & According to the 1994 film Forrest Gump, what was Gump's IQ? & 75 & $\bm{(50, 110)}$\\
  5 & In years, how long did it take Michelangelo to paint the ceiling of the Sistine Chapel? & 4 & $(1, 3)$\\
  6 & In months, how long did it take Apple to sell 100 million iPods? & 66 & $(80,120)$\\
  7 & How many episodes aired of the TV sitcom Friends, not including specials? & 236 & $(250, 300)$\\
  8 & In what year was the first modern Summer Olympics held? & 1896 & $\bm{(1896, 1900)}$\\
  9 & In miles, how long is the main wall of the Great Wall of China, excluding all of its secondary branches? & 4163 & $\bm{(1000, 8000)}$\\
  10 & How many states were part of the United States in 1860? & 33 & $\bm{(0, 50)}$\\
  \bottomrule
\end{tabular}

\end{center}
Total score: 6 out of 10

\subsection*{Sample Visualization} 
\label{sub:sample_visualizations}
\begin{center}
    \includegraphics[width=0.5\textwidth]{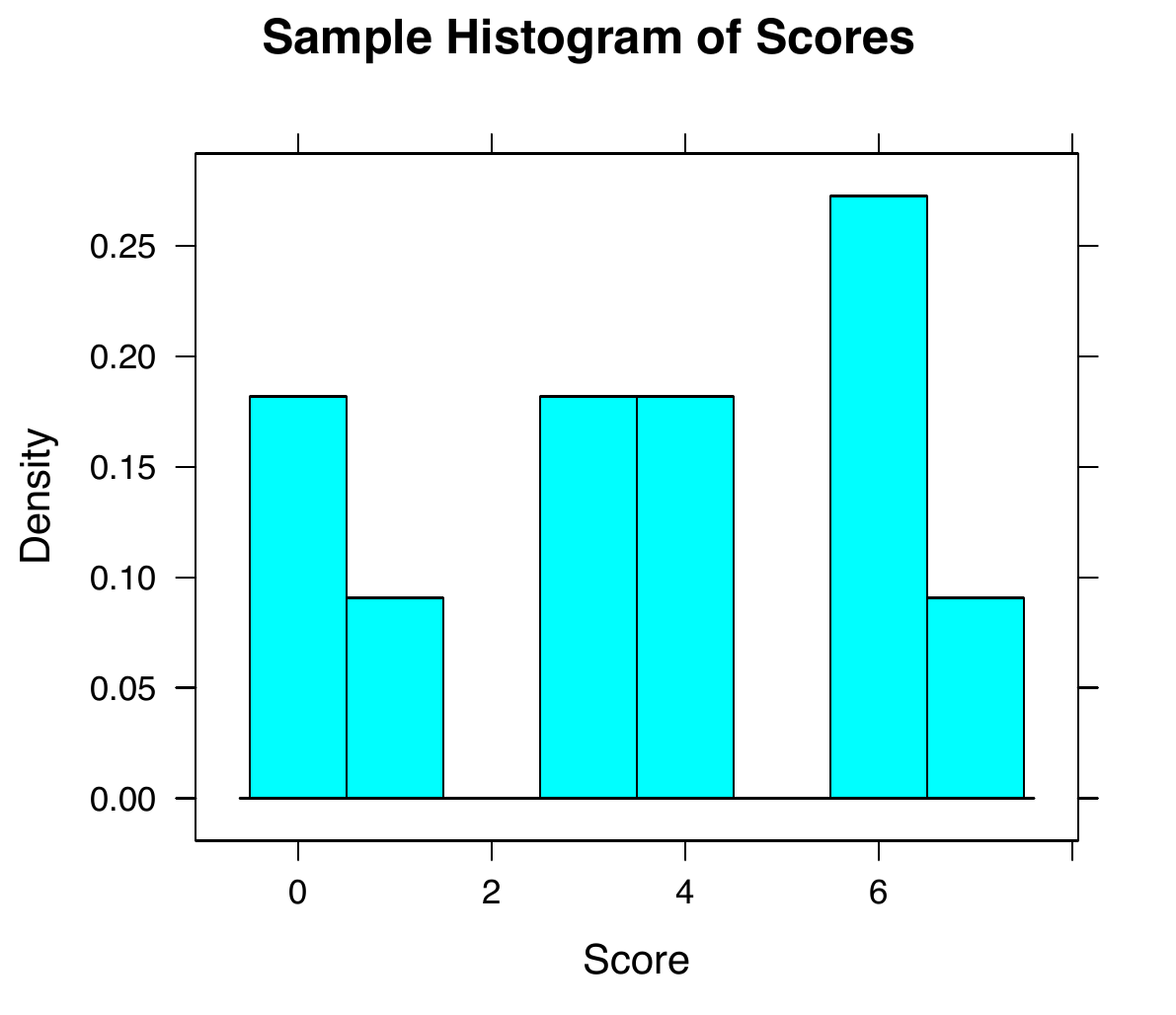}

\end{center}

\newpage
\section*{CAOS Test Results}
In the graduate-level course where the activity was repeatedly used, the Comprehensive Assessment of Outcomes in a First Statistics course (CAOS) test, was used to evaluate overall conceptual understanding of statistics. The CAOS pre- and post-tests were administered at the beginning and end of the semester, respectively. We specifically evaluated the student responses to four questions (questions 28 through 31) on the CAOS pre- and post-tests that evaluate understanding about confidence intervals. The percentage of correct responses on all 4 questions increased from 61.0\% on the pre-test to 76.7\% on the post-test. Students showed larger improvements on Questions 28 and 29 than on Questions 30 and 31. 

\begin{table}[ht]
\centering
\begin{tabular}{cccc}
  \toprule
  & \multicolumn{2}{c}{\% correct}  \\
  question & pre-test & post-test \\
  \midrule
  28 & 37.5 & 93.3 \\
  29 & 68.8 & 80 \\
  30 & 43.8 & 46.7 \\
  31 & 93.8 & 86.7 \\
  average & 60.975 & 76.675 \\
   \bottomrule
\end{tabular} 
\caption{Summary statistics from the CAOS pre-test ($N$=16) and post-test ($N$=15) questions 28 through 31, which directly pertain to confidence intervals.}\label{tbl:caos}
\end{table}
\end{document}